# Tightly confining lithium niobate photonic integrated circuits and lasers

Zihan Li,[1, *] Rui Ning Wang,[1, *] Grigorii Lihachev,[1, *] Zelin Tan,[1] Viacheslav Snigirev,[1] Mikhail Churaev,[1] Nikolai Kuznetsov,[1] Anat Siddharth,[1] Mohammad J. Bereyhi,[1] Johann Riemensberger,[1, †] and Tobias J. Kippenberg[1, 2, ‡]

[1]Institute of Physics, Swiss Federal Institute of Technology Lausanne (EPFL), CH-1015 Lausanne, Switzerland
[2]Center of Quantum Science and Engineering (EPFL), CH-1015 Lausanne, Switzerland

**Photonic integrated circuits are indispensible for data transmission within modern datacenters and pervade into multiple application spheres traditionally limited for bulk optics, such as LiDAR and biosensing [1]. New applications and higher performance are enabled by the diversification of optical waveguide materials past silicon-on-insulator. Of particular interest are ferroelectrics such as Lithium Niobate, which exhibit a large electro-optical Pockels effect enabling ultrafast and efficient modulation, but are difficult to process via dry etching [2]. For this reason, etching tightly confining waveguides - routinely achieved in silicon or silicon nitride - has not been possible. Diamond-like carbon (DLC) was discovered in the 1950s [3] and is a material that exhibits an amorphous phase, excellent hardness, and the ability to be deposited in nanometric thin films. Its use today is pervasive, ranging from applications for hard disk surfaces [4] and medical devices [5] to low friction coatings for automotive components [6]. It has excellent thermal, mechanical, and electrical properties, making it an ideal protective coating. Here we demonstrate that DLC is also a superior material for the manufacturing of next-generation photonic integrated circuits based on ferroelectrics, specifically Lithium Niobate on insulator (LNOI). Using DLC as a hard mask, we demonstrate the fabrication of deeply etched, tightly confining, low loss photonic integrated circuits with losses as low as 5.6 dB/m. In contrast to widely employed ridge waveguides [7, 8], this approach benefits from a more than 1 order of magnitude higher area integration density while maintaining efficient electro-optical modulation, low loss, and offering a route for efficient optical fiber interfaces. As a proof of concept, we demonstrate a frequency agile hybrid integrated III-V Lithium Niobate based laser with kHz linewidth and tuning rate of 0.7 Peta-Hertz per second with excellent linearity and CMOS-compatible driving voltage. Our approach can herald a new generation of tightly confining ferroelectric photonic integrated circuits, in particular for applications in coherent laser based ranging [9] and beamforming [10], optical communications [7], and classical [11] and quantum computing networks [12].**

Photonic integrated circuits based on silicon (Si) have transitioned from academic research to use in data centers over the past two decades [1]. In the second wave of development, silicon nitride has emerged as an integrated photonics platform [13], offering lower loss, nonlinear operation, high power handling capability, and a wide optical transparency window. New capabilities include chipscale optical frequency comb sources[14], traveling-wave optical parametric amplifiers [15, 16] as well as integrated lasers that operate in the visible spectral range Tobias [17], and that rival the phase noise of fiber lasers[18] and offer unprecedented fast tuning capability using MEMS integration [9], with applications ranging from LiDAR [19], coherent communications [20] to visible wavelength lasers for atomic based quantum sensors and clocks [17]. The advent of Lithium Niobate on insulator [21, 22] - and ferroelectric thin film materials on insulator ('OI') in general - can extend the functionality further by offering one of the highest Pockels coefficients, required to realize volt level high speed modulators [7, 8], electro-optical frequency combs[23], or photonic switching networks [24] or delay lines [25], and on-chip broadband spectrometers [26, 27]. Periodic poling of thin film-based $LiNbO_3$ ridge waveguides has allowed on-chip frequency doublers [28, 29], squeezed light sources[30], and optical parametric oscillators [31]. $LiNbO_3$ also features large second-order nonlinear susceptibility for optical frequency conversion [32] and piezoelectric coefficient enabling advanced on-chip acousto-optics [33]. Integrated photonic circuits critically rely on the achieving wafer scale manufacturing, which exhibits low loss and attains lithographic precision and reproducibility. A critical manufacturing step is the etching process, which transfers the lithographic pattern into the photonic device. While for currently employed foundry compatible photonic material platforms, in particular, silicon or silicon nitride, mature processing is available, the latter cannot be readily extended to the rapidly emerging platforms based on ferroelectric materials, such as Lithium Niobate on insulator [2]. Although major progress has been made, currently employed LNOI based devices are compounded by insufficient etch mask selectivity, typically based on silicon or silica. Hence, state of the art in $LiNbO_3$ integrated photonics technology [34] are shallow ridge waveguides with close to unity ridge to slab ratios and strongly slanted sidewalls ($\approx 60°$) that are manufactured using reactive ion beam etching and $SiO_2$ based etching masks. The etching depth and sidewall angles are limited by the erosion of the soft $SiO_2$ mask during the physical etching. The thick slab entails strong limitations in terms of waveguide bending radius as well as requiring more challenging process control due to the partial etch. This complicates efforts to establish and qualify process design kits (PDK) for $LiNbO_3$ photonic integrated waveguide devices. Direct etching of lithium niobate is usually based on argon ion bombardment, which is a strong physical process that cannot achieve a high etch selectivity between lithium niobate and common mask materials, such as $SiO_2$ and a-Si. Therefore, soft silicon-based masks lead to shallow sidewall angles and are not suitable for deep etching. Metallic masks, such as chromium [35], feature high etch selectivity but are polycrystalline with rough sidewalls and optically absorbent etch product redeposition. Heterogeneous integration of $LiNbO_3$ with silicon or silicon nitride integrated photonic circuits is another approach. Such hybrid waveguides have been successfully achieved[36, 37], yet exhibit additional complexity due to the use of wafer bonding onto pre-fabricated substrates, and require mitigating losses from transitioning into and out of the bonded areas. Competing hybrid waveguide technologies are primarily reliant on high loss materials such as a-Si:H [38] and non-stochiometric silicon nitride [39] or require precise wafer bonding on premanufactured waveguide substrates [36, 37]. Beyond applications for electro-optic modulators, an integrated photonics platform with an electro-optic capa-bility and low propagation loss allows the realization of integrated frequency-agile laser sources, which feature very fast and linear mode-hop-free frequency tuning, and can be realized via self-injection locking of a III-V gain media to a high-$Q$ optical cavity. This technique has already enabled hybrid integrated lasers with sub-Hz Lorentzian linewidth [40] using microresonators with low confining $Si_3N_4$ waveguides [41]. Hence a $LiNbO_3$ photonic integrated circuit platform featuring ultra-low loss fully etched strip waveguides with vertical sidewalls, already well established and successfully commercially employed for silicon and silicon nitride, would be highly desirable. Here we overcome this challenge and demonstrate a low loss, tightly confining $LiNbO_3$ PIC platform based on a novel etching processing technique. We further demonstrate high-$Q$ microresonators with an intrinsic linewidth of 27 MHz and demonstrate a hybrid integrated frequency-agile laser with fast tuning rate up to 1 MHz and intrinsic linewidth of 6.5 kHz. In contrast to previously demonstrated narrow linewidth frequency agile lasers, the approach using $LiNbO_3$ based PIC enables to lower the drive voltage to CMOS level ($< 2$ Volts). Central to our results is the establishment of a novel hard mask process based on diamond like carbon (DLC).

---
* These authors contributed equally to this work.
† johann.riemensberger@epfl.ch
‡ tobias.kippenberg@epfl.ch

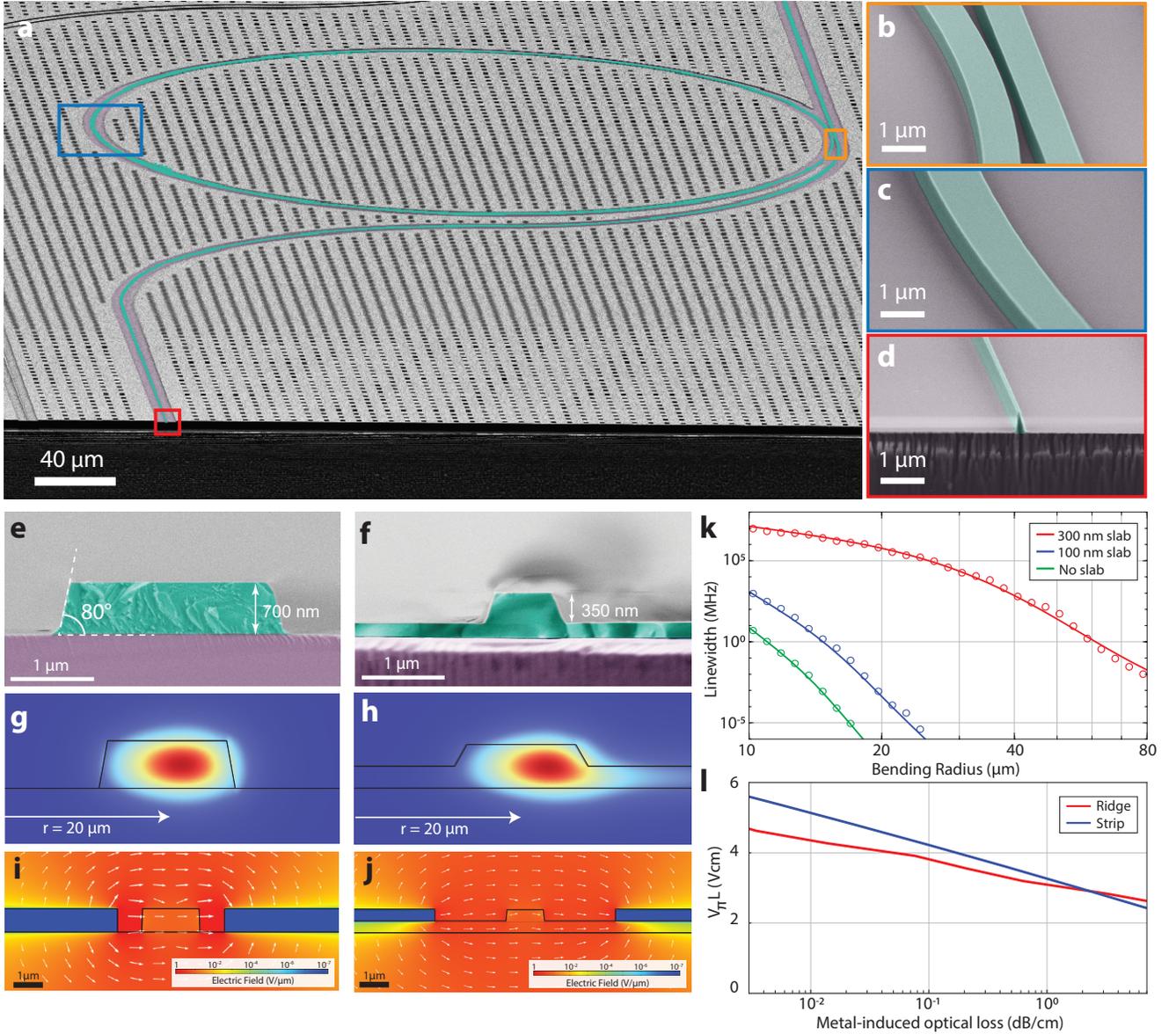

**Figure 1. Tightly confining LNOI platform. (a)** Scanning electron micrograph (SEM) of a $LiNbO_3$ -based photonic integrated circuit with high resolution insets of microring coupling section (**(b)**,orange), curved waveguide (**(c)**,blue), and inverse taper for fiber coupling (**(d)**,red). **(e)** SEM of a fully etched optical waveguide with cross section $2\,\mu m \times 0.7\,\mu m$. **(f)** SEM of a partially etched ridge waveguide $1.5\,\mu m \times 0.6\,\mu m$, $350\,nm$ etch depth and $250\,nm$ $LiNbO_3$ slab. (g,h) Optical mode field distribution of curved waveguides corresponding to (e,f) with $20\,\mu m$ bend radius. **(i,j)** Electrical field distribution in log-scale for slab and ridge waveguides with electrode spacing $0.75\,\mu m$ and $2.5\,\mu m$, which are selected to feature the equal metal-induced optical losses. **(k)** FEM simulations of LN waveguide bending loss for different slab heights and waveguide cross-section $2\,\mu m \times 0.6\,\mu m$. **(l)** FEM Simulation of the voltage-length product of electro-optical modulation for strip and ridge waveguides as a function of metal-induced waveguide loss.

DLC was discovered in 1953 [3] as byproduct of a carburization experiment by H.P.P. Schmellenmeier, who obtained a black bright film on a metal alloy in a glow-discharge plasma of $C_2H_2$, which showed scratch resistance and amorphous characteristics in X-ray scattering. At the beginning, this novel material was used for protective coating [6, 42], data storage [4] and sensors [5, 43] due to its excellent electrical and mechanical properties. DLC films can be deposited by plasma enhanced chemical vapor deposition (PECVD) with hardness up to 20 GPa or physical vapor deposition (PVD) with hardness up to 80 GPa [44]. As an amorphous carbon allotrope, DLC presents a smooth boundary after etching in oxygen plasma. Hirakuri et. al. applied DLC films as etch masks for integrated circuit fabrication in 1999 [45]. The low argon ion sputtering yield [46] and excellent chemical stability of DLC are much desired for the etching of all integrated optical materials etching such as $LiNbO_3$, $SiO_2$, $BaTiO_3$, 0 and $Si_3N_4$ as they limit mask erosion processes that determine surface roughness and lead to slanted waveguide sidewall angles.

**Tightly confining LNOI photonic integrated circuit platform:** Figure 1(a) depicts a scanning electron micrograph (SEM) of an optical microresonator manufactured with our DLC-based process. Colored insets highlight the waveguide sidewalls and high aspect ratio positive and negative features such as directional couplers and inverse tapers (see Fig. 1(b,c,d)). The critical distance is 300 nm for positive and 200 nm for negative features with a sidewall angle of 80°. Inverse tapers with 250 nm width expand the waveguide mode sufficiently to support low loss input coupling from a lensed fiber to the $LiNbO_3$ PIC, reaching 3 dB/facet level without the need for a two step etching process that is required to reach similar levels in ridge waveguides [47] and and heterogeneously integrated hybrid $LiNbO_3$ -$Si_3N_4$ waveguides [36, 37] to mitigate the typical 5-10 dB facet inser-tion loss. Figure 1(e,f) compares our waveguides with a ridge $LiNbO_3$ waveguide with 350 nm etch depth and 250 nm remaining slab with a top width of 1.5 μm where low-loss operation has been demonstrated [48]. Crucially, the $LiNbO_3$ slab induces a continuum of leaky modes with a high refractive index that couple to the guided mode in curved waveguide sections. The numerical simulations presented in Fig. 1(g,h,k) reveal a 9 and 12 order of magnitude reduction of bending losses for curved waveguides with 20 μm radii if the slab is thinned to 100 nm and fully removed, respectively. Similar top waveguide widths of 2 μm and thicknesses of 0.7 μm are assumed in the numerical simulation. Fully etched strip waveguides (see Fig. 1(a)) feature tight optical mode confinement at bending radii below 20 μm, enabling dense integration of optical components on chip with a factor of 16 area density improvement over currently employed ridge waveguide structures. While a residual slab layer can improve the modulation efficiency due to the high dielectric constant $LiNbO_3$ , the increased optical confinement allows to place electrodes closer to the waveguide, which enables electro-optical modulation efficiencies that are competitive with ridge waveguides [7] (see SI) and hybrid waveguides [8, 37] without inducing excess Ohmic losses at the metal electrode interfaces (see Fig. 1(i,j)). Indeed our numerical simulations show that only a 10% penalty in DC modulation voltage-length product is incurred by transitioning to a fully etched system at an Ohmic loss target of $0.1\,dB\,m^{-1}$ (see Fig. 1(l)).

**Fabrication process with DLC hard mask:** Figure 2(a) depicts the schematic process flow of our low-loss $LiNbO_3$ strip waveguide fabrication. We start the fabrication using commercially available thin film lithium niobate on insulator wafer (NanoLN) with X-cut 700 nm thick $LiNbO_3$ layer, 4.7 μm buried oxide (thermally grown) on and 525 μm thick Si substrate. The salient feature of our fabrication process is the use of a 300 nm thick DLC film, which is grown via plasma-enhanced chemical vapor deposition (PECVD) from a methane precursor, as a hard mask material for the physical ion beam etching process. The diamond hardness of the as-deposited DLC film can achieve 19-23MPa (see Fig. 2(f) and Methods), which is up to two times harder than $SiO_2$ and $LiNbO_3$ . The chemical composition of the film is measured via Raman spectroscopy [44] (see Fig. 2(h) and Methods) . Analyzing the two C-C stretch modes (D,G) around $1300\,cm^{-1}$ and $1550\,cm^{-1}$, we find that our film contains up to 10% sp3-hybridized C-atoms and a large hydrogen content, which classifies the material as amorphous diamond like carbon (a-C:H), typical of PECVD deposition using methane precursor. We have carefully optimized the process to facilitate wafer scale uniform growth of DLC for minimal stress and particle contamination, which is pivotal to the manufacture of photonic integrated waveguide structures. The optical waveguide pattern is structured by DUV stepper photolithography (248 nm) and first transferred into a $Si_3N_4$ mask layer using standard fluoride-chemistry based plasma etching. This intermediate mask has excellent resistance to oxygen plasma etching, which we use to transfer the waveguide pattern into the DLC hard mask (see Fig. 2(c)). With optimized argon ion beam etching (IBE), the etching selectivity between LN and DLC is up to ×3, which enables deep etching and steep sidewalls. Figure 1(e) contains a SEM cross section of the fully etched $LiNbO_3$ strip waveguide with waveguide base width of $3.3\,\mu m$ and height of $0.7\,\mu m$ featuring a 80° sidewall angle. Currently employed soft $SiO_2$-based masks are limited an etch selectivity of ×1 in comparison. The film hardness directly correlates with the material etching rate in IBE, which is plotted as the etch selectivity compared to $LiNbO_3$ in Fig. 2(g). This is remarkable in particular because our film hardness lies at the lower limit of the attainable hardness range of DLC thin films[44]. We remove the rough $LiNbO_3$ redeposition on the waveguide sidewall using SC-1 solution ($NH_4OH:H_2O_2:H_2O=1:1:5$) (see Fig. 2(d)) and the residual DLC mask (see Fig. 2(d)) using oxygen plasma to reveal the waveguide core (see Fig. 2(e)). Next, we fabricate the electrodes via a DUV-stepper lithography based lift-off process for which we deposit $5\,nm$ titanium adhesion layer and a $400\,nm$ gold layer via electron-beam evaporation. Finally, the wafer is separated into chips via deep dry etching followed by backside grinding to obtain clean, vertical and smooth facets without a silicon pedestal for efficient edge coupling. Figure 2(j) depicts a normalized transmission (blue) and Lorentzian fit (red) of a microring resonance with $100\,GHz$ FSR fabricated using our process and measured using frequency-comb-assisted diode laser spectroscopy [49]. The linewidth and dispersion measurement of the resonance at $193.57\,THz$ yields an intrinsic photon loss rate $\kappa_0/2\pi = 27.1\,MHz$, which corresponds to a quality factor of greater than 7 million and a linear propagation loss $\alpha = n_g/c \cdot \kappa_0$ of $5.6\,dB\,m^{-1}$, with the group velocity index $n_g = 2.275$ of our $LiNbO_3$ strip waveguide. Figure 2(l) the most probable value $\kappa_0/2\pi$ of the 9 chips at different positions on the wafer layout with a most probable intrinsic loss photon rate of $72\,MHz$ in the worst wafer field, which corresponds to a linear propagation loss of $15\,dB\,m^{-1}$. The reticle design contains 16 5x5 mm chips in each of 9 fields uniformly exposed over a 4-inch wafer. Figure 2(k) depicts the measured integrated dispersion $D_{int}/2\pi = \omega(\mu) - \omega_0 - D_1/2\pi \cdot \mu$ of the LN microresonator with FSR $D_1/2\pi = 17.9\,GHz$ featuring anomalous microresonator dispersion of $D_2/2\pi = 20.9\,kHz$, corresponding to a group velocity dispersion (GVD) of $\beta_2 = -66\,fs^2\,mm^{-1}$. We observe mode mixing at wavelengths shorter than $1580\,nm$, which we attribute to the mixing of fundamental TE and TM modes in the ring resonator due to the inherent birefringence of the material.

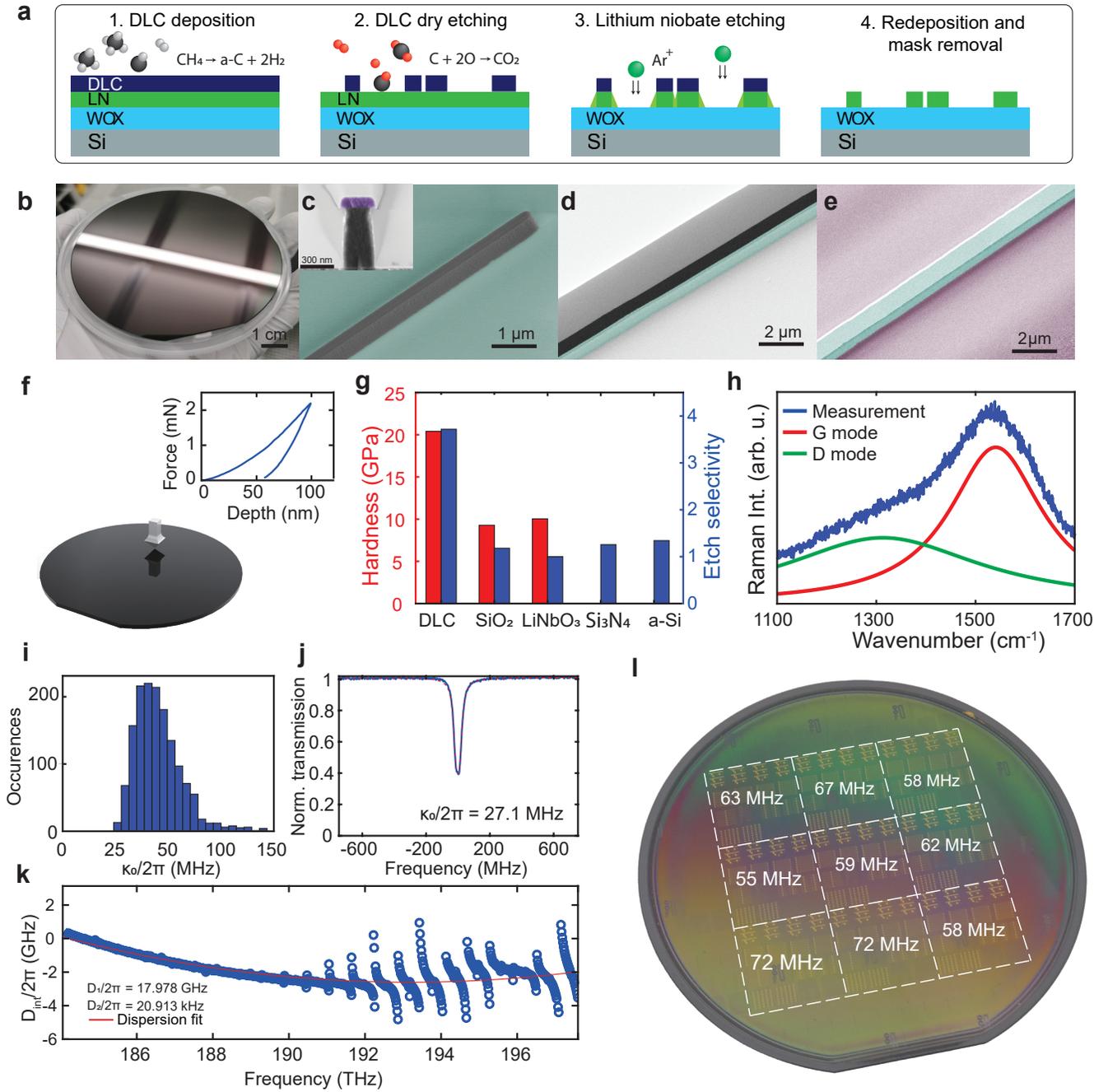

**Figure 2. Device fabrication process and characterization. (a)** Process flow of the LNOI waveguide fabrication including diamond-like carbon (DLC) hard mask deposition via plasma enhanced chemical vapor deposition (PECVD) from methane precursor, DLC dry etching via oxygen plasma, lithium niobate etching via argon ion beam etching (IBE), followed by redeposition and mask removal. LiNbO$_3$ is illustrated in green, SiO$_2$ in light blue, DLC in dark blue and Si in grey. **(b)** Photo of a deposited DLC film on the silicon wafer. SEM images of the taper pattern and the cross section with 250 nm width after DLC dry etching **(c)** DLC mask and the LiNbO$_3$ waveguide after redeposition cleaning **(d)** and the LN waveguide with 0.7 μm height and 2 μm width after mask removal **(e)**. In the SEM images, the LiNbO$_3$ is colored in green, DLC in black, Si$_3$N$_4$ in purple and SiO$_2$ in light purple. **(f)** The schematic and result of indentation hardness measurement for DLC. **(g)** Hardness and etching selectivity chart for different materials. **(h)** Raman spectrum and fitting result for DLC film. **(i)** Histogram of TE00 resonances from a single microring LNOI resonator showing the most probable value of $\kappa_0/2\pi = 40$ MHz. **(j)** Microresonator transmission (blue) and fit (red). Linewidth measurement of the resonance at 193.57 THz using frequency-comb-assisted diode laser spectroscopy results in $\kappa_0/2\pi = 27.1$ MHz. **(k)** Measured integrated dispersion of the LN microresonator with free spectral range 17.9 GHz and anomalous dispersion of $D_2/2\pi = 21$ kHz. Avoided mode crossings due to TE-TM mode mixing are visible in the 190 − 198 THz frequency range. **(l)** The most probable value $\kappa_0/2\pi$ of the 9 chips at different positions on the wafer layout. The reticle design contains 16 chips in each of 9 fields uniformly exposed over a 4-inch wafer.

**Ultrafast tunable low noise hybrid integrated III-V/LNOI laser:** To demonstrate the utility of the platform, we demonstrate a ultrafast tunable low noise laser, based on hybrid integration of a hybrid integrated III-V/LNOI laser. We exploit the laser self-injection locking effect by directly edge-coupling an InP distributed-feedback (DFB) laser with a $LiNbO_3$ racetrack resonator. The racetrack resonator used in this experiment features the free-spectral range (FSR) of $81\,GHz$, a waveguide width of $2\,\mu m$, and a waveguide height of $0.7\,\mu m$. The minimum bending radius in the 180° bends is $100\,\mu m$ and the straight sections are $400\,\mu m$ long. We couple light into and out of the racetrack resonator using a 5° Pulley coupler [50] and a bus waveguide width of $1.25\,\mu m$ for phase matching. Straight and curved waveguide sections are connected by C4-continuous spline bends to avoid excitation of high order modes [51]. The intrinsic cavity linewidth of the resonance used is $\kappa_0/2\pi = 130\,MHz$, and the power reflection coefficient on resonance is $R = 0.07$. The DFB laser diode is mounted on a 3D translation stage, and laser self-injection locking is initiated by edge-coupling DFB laser to the $LiNbO_3$ chip (see Fig. 3(a)) and tuning the laser current and temperature to match the resonance frequency of the $LiNbO_3$ cavity. The bulk and surface scattering [52] inside the $LiNbO_3$ waveguide induces back-scattered light, provides spectrally narrowband fast optical feedback and triggers laser self-injection locking. This results in laser frequency noise reduction (the ratio of the free-running laser linewidth to the linewidth of the self-injection-locked laser), which is quadratically proportional to the loaded $Q$ factor of the $LiNbO_3$ cavity resonance [53]. The gap between the DFB and the $LiNbO_3$ chip is adjusted for the optimal optical feedback phase, which leads to the maximum self-injection locking range. The free space output power of the DFB is $62\,mW$ at $238\,mA$ current. The output power of the hybrid integrated III-V/LNOI laser measured in the output fiber is $1\,mW$ operating at a lasing wavelength of $1556.3\,nm$ (see Fig. 3(b)), resulting in a total power loss of $17.9\,dB$ loss from the DFB laser, through the LN chip-lensed fiber including optical power dissipation in the microresonator. This can further be improved by optimizing the taper design on the laser interface to match the output mode of the DFB [54] using a horn (outverse) taper. We employ heterodyne beatnote spectroscopy with an auxiliary fixed frequency laser and measure the frequency tuning curve of the hybrid integrated $LiNbO_3$ laser. Figure 3(d) shows a spectrogram of hybrid laser frequency change upon the linear tuning of the diode current. The areas between dashed lines correspond to the regions where the DFB laser is self-injection locked with minimal lasing frequency change during the laser current scan with characteristic laser frequency abruptly jumps in and out of the locked state at $8\,s$ and $11\,s$ as the laser current is increased and at $14\,s$ and $17\,s$ as the laser current is decreased, respectively. The estimated full self-injection locking range of the frequency detuning of the free-running laser relative to the resonance frequency is $2.5\,GHz$. The heterodyne beatnote measurement also allows to measure laser frequency noise using a fast photodetector (FPD) and an electrical spectrum analyzer (ESA). Figure 3(c) shows that the frequency noise of the hybrid integrated self-injection locked laser is suppressed by more than 20 dB across the entire spectrum (see the full plot in the SI). The laser frequency noise reaches a plateau (white noise floor) of $2000\,Hz^2\,Hz^{-1}$ at $5\,MHz$ offset. We attribute the broad noise peak at $1\,MHz$ offset to DFB noise and not as a result of the self-injection locking to the $LiNbO_3$ resonator. By integrating frequency noise up to the frequency of the interception point with a line $S_\nu(f) = 8 \cdot \ln 2f/\pi^2$ ($72\,kHz$ for our device), we obtain a frequency measure $A$ for the full width at half maximum of the linewidth $(8 \cdot \ln 2A)^{1/2}$ [55]. The FWHM linewidth is $907\,kHz$ at $1\,ms$ integration time, $2.9\,MHz$ at $10\,ms$, and $3.5\,MHz$ at $100\,ms$.

**Frequency-agile LNOI-based laser tuning:** Next, we demonstrate the ability of ultra-fast frequency actuation of the laser using the Pockels effect of the $LiNbO_3$ microresonator. The frequency spectrum of the electro-optical response of the $LiNbO_3$ microresonators is measured using a vector network analyzer (see Fig. 4(a)). A continuous-wave laser at 1550 nm is coupled into the device using a lensed fiber and the laser frequency is tuned to the slope of the $LiNbO_3$ cavity resonance. A swept radiofrequency signal of -5 dBm power is applied from port 1 of the network analyzer to the electrodes of the device using RF probes, and the transmitted light intensity modulation is detected by a 12 GHz photodiode (New Focus 1544), which is sent back to port 2 of the network analyzer. We next demonstrate the frequency agility of our hybrid laser by setting the diode current to the center of the locking plateau and applying a triangular voltage to electrodes. Changes in the $LiNbO_3$ cavity resonance frequency via Pockels effect will maintain self-injection locked laser operation and therefore lead to frequency tuning. Thus, the applied voltage to $LiNbO_3$ device is transduced directly to changes in the laser frequency. Figure 3(d) shows the maximal range over which the $LiNbO_3$ resonator frequency can be tuned while maintaining laser locked to the $LiNbO_3$ cavity. Figure 4(b) reveals one of the advantages of the $LiNbO_3$ platform in comparison to piezoelectric actuators - the modulation response function is flat up to the cavity cutoff frequency and no mechanical modes of the chip are excited [9]. In the self-injection locking range, the change of $LiNbO_3$ microresonator frequency directly changes the laser output frequency without additional feedback on the diode current. Figure 4(c) shows the main results of the heterodyne beat experiment with the DFB laser locked to a $LiNbO_3$ racetrack resonator. We define chirp nonlinearity as the root mean square (RMS) deviation of the measured frequency tuning curve from a perfect triangular ramp that is determined with least-squares fitting. Figure 4(c) presents the processed laser frequency spectrograms and the corresponding RMS nonlinearities upon applying to the electrodes triangular ramps with $2\,V_{p-p}$ amplitude at 10 kHz, 100 kHz and 1 MHz frequencies. Measured tuning efficiency is $380\,MHz\,V^{-1}$ using only a single electrode on the side of the 81 GHz FSR racetrack resonator for self-injection locking. At 1 MHz modulation frequency, the achieved RMS nonlinearity is as low as 3.9 MHz (relative nonlinearity 0.5%), which surpasses benchtop laser systems. The tuning range of 760 MHz at high ramping speeds from 10 kHz up to 1 MHz, with small chirping RMS nonlinearities from 0.3% to 0.5% proves the frequency agility of our hybrid integrated laser. We also do not observe any tuning hysteresis or inherent nonlinear response of the $LiNbO_3$ microresonator.

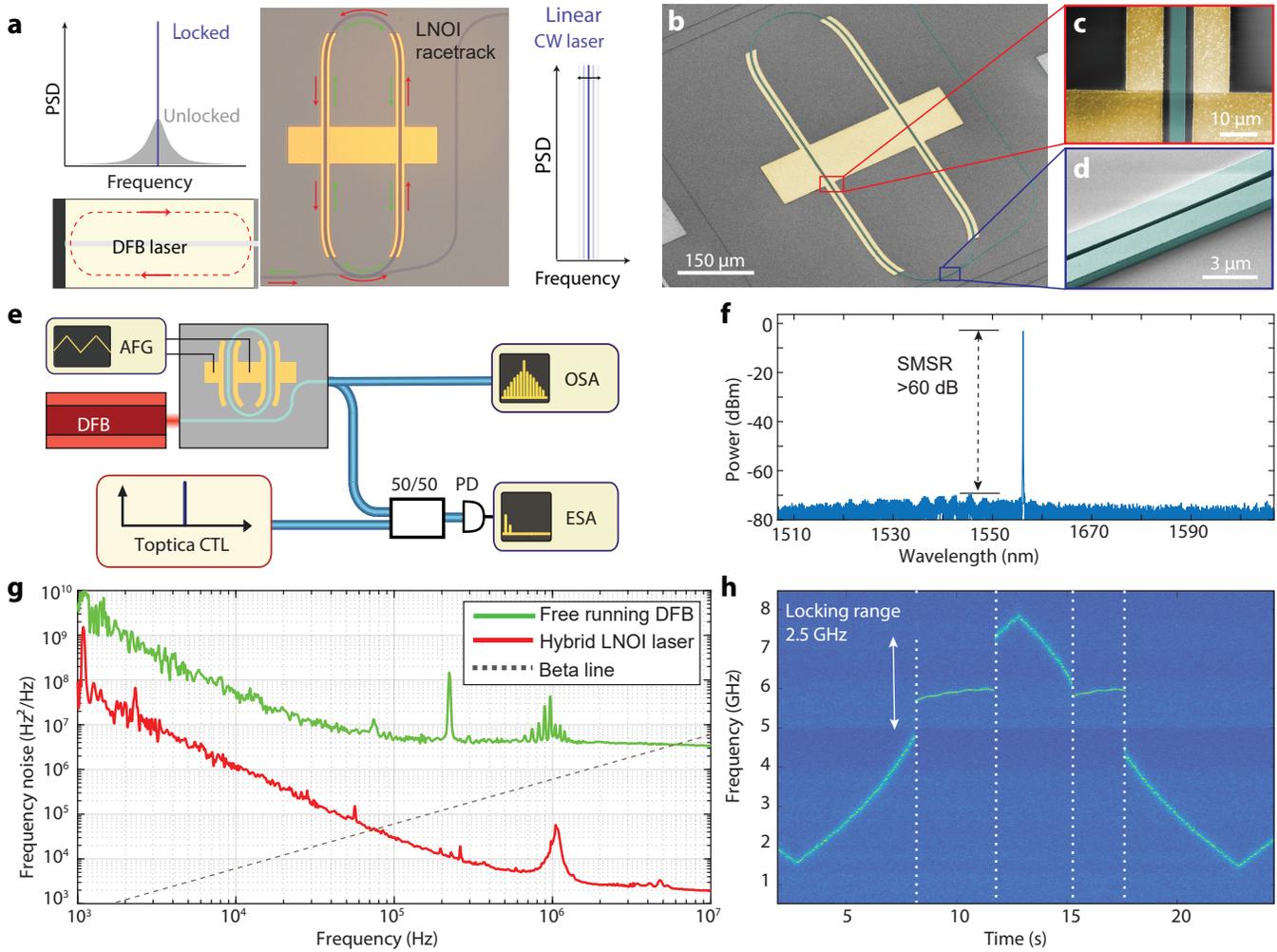

**Figure 3**. **Hybrid LNOI laser performance characterization.** **(a)** Schematic of hybrid LNOI laser. DFB laser is self-injection-locked to an LNOI photonic chip-based racetrack resonator. The bulk and surface scattering inside the LNOI waveguide induces back-scattered light and triggers the laser self-injection locking and frequency noise reduction. **(b)** SEM of the racetrack resonator with electrodes including high resolution insets of the waveguide with electrodes **(c,**red) and the coupling section **(d,**blue) **(e)** Schematic of the optical setup for SIL laser characterization. Heterodyne beatnote measurement using free running Toptica CTL as a reference allows to measure laser frequency noise using fast photodetector (FPD) and electrical spectrum analyzer (ESA). Optical spectrum analyzer (OSA). **(f)** Optical spectrum of the hybrid LNOI laser emission. The side mode suppression ratio (SMSR) is greater than 60 dB. **(g)** Single-sided PSD of frequency noise of the hybrid LNOI laser upon self-injection locking to racetrack resonator with 81 GHz FSR (red) and in free-running regime (green). **(h)** Spectrogram showing laser frequency change upon the linear tuning of the diode current, dashed areas correspond to the 2.5 GHz range where the laser is self-injection locked with minimal lasing frequency fluctuations and reduced linewidth.

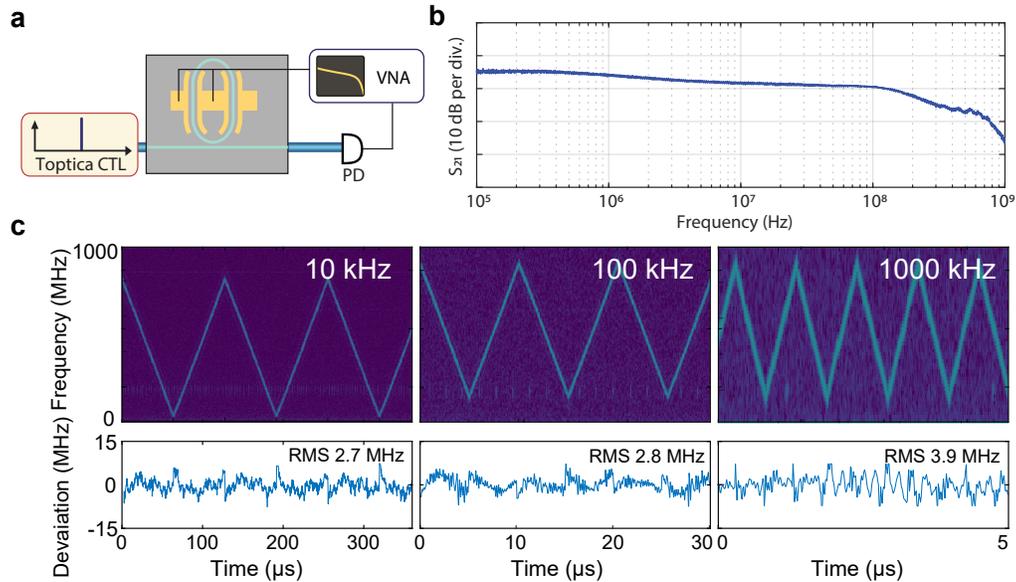

**Figure 4**. **Fast tuning of the hybrid LNOI laser.** **(a)** Experimental setup for electro-optical response characterisation of the LNOI microresonator laser using an external laser (Toptica CTL), a photoreceiver (PD) and a vector network analyzer (VNA). The beatnote is recorded on a fast oscilloscope (DSO) and analysed with short-time Fourier transforms. **(b)** Measured responses of the electro-optic actuation for the 81 GHz FSR microresonator showing flat actuation bandwidth up to the cavity cutoff frequency. **(c)** Time-frequency spectrogram of the heterodyne beatnotes for triangular chirp repetition frequencies from 10 kHz to 1 MHz. The frequency excursion is 761 MHz at 10 kHz tuning rate, 730 MHz at 100 kHz and 721 MHz at 1 MHz chirp rate. Bottom row: residual of least-squares fitting of the time-frequency traces with symmetric triangular chirp pattern.

**Discussion & Conclusion:** We have presented a novel platform for lithium niobate on insulator integrated photonics based on a deeply etched strip waveguide with tight optical confinement based on a novel microstructuring process featuring amorphous carbon films as etch mask. Our process is readily applied for a wide variety of photonic materials that are resistant to oxygen plasma etching, which includes all silicon, its nitrides and oxides, and notably also ferroelectric oxides beyond $LiNbO_3$ that have attained large interest recently. At similar loss levels compared to $LiNbO_3$ -based ridge waveguides, our fully etched geometry affords a 4 times smaller minimum bend radius, which corresponds to a potential increase of photonic component density by a factor of 16 without sacrificing significant electro-optical modulation efficiency. Second, ridge waveguides are less suited for precise group velocity dispersion (GVD) engineering due to the sidewall angle and geometry variation, and the presence of the ridge limits the range of anomalous dispersion attainable. Hence, the use of high-confinement and fully etched strip waveguides unlocks not only the wafer scale high yield manufacturing of efficient optical modulators [7] but also of photonic integrated circuits with dense component placement, up to a level achieved previously only in Si and $Si_3N_4$ -based integrated photonics, which important for application that require a very high component count or small die size. The latter is e.g. key for the application of thin film $LiNbO_3$ in photonic quantum or photonic neuromorphic computing[11], where large networks of passive and tunable photonic components are required [12]. In a similar manner, the increased optical confinement can greatly benefit other electro-optical quantum technologies such as quantum coherent microwave to optical converters through the decreased optical cavity mode volume and increased single photon coupling efficiency [56, 57]. We use this platform to demonstrate a $0.7\,PHz\,s^{-1}$ tunable laser via self-injection locking with 0.5% RMS nonlinearity up to a fundamental chirp modulation frequency of 1 MHz using purely triangular drive signals with no signal preconditioning. Compared to a very recent demonstration of a III-V/lithium niobate laser using a Vernier filter-based scheme and transfer printing [58], the self-injection locking technique provides substantially improved tuning linearity at high tuning speed, which is relevant to applications in coherent laser ranging and fiber sensing. In contrast to thermal or piezoelectric [59–61] tuning of optical microresonators, which requires engineering of the phononic mode spectrum of the chip [9], electro-optic materials can enable flat actuation bandwidth and very fast optical tuning out to multi-MHz bandwidths. Such a system is an attractive candidate for applications in coherent (FMCW) laser ranging and critically enables to reduce the voltage driving to CMOS level, while achieving GHz tuning range, excellent linearity, no measureable hysteresis, and low laser phase noise. Moreover, we note that the incorporation of erbium and other rare earth ions into $LiNbO_3$ thin films is possible [62] and enables to build active photonic devices such as on-chip microdisk C-band lasers [63] and on-chip quantum memories [64, 65].

**Methods:**

**DLC hard mask deposition and characterization:** Ion beam etching requires mask materials with high hardness and resistance to ion bombardment. We deposit DLC film in a plasma etcher (OXFORD). The chemical properties of PECVD DLC were characterized by Raman spectroscopy with 532 nm excitation (RENISHAW inVia Raman Microscope). Two Raman-active C-C stretch vibration modes are observed, the D-mode (disorder), which corresponds to sp2-hybridized C-atoms organized in linear chains and broken rings, and the G-mode (graphene), which corresponds to the stretch vibration of C6 rings. The C-C stretch vibration of sp3-hybridized atoms is not observed because it is at least 50 times less Raman active than the sp2-hybridized modes for visible wavelength excitation and does overlap with the D-mode. We analyze the Raman spectrum by fitting the sum of a Lorentzian and a skewed Lorentzian to the D- and G-modes, respectively [44]. The chemical composition of the material and hybridization ratios can be inferred from the center frequency of the G-mode and the D/G ratio and we extract a sp3 ratio of 10%. The hardness of DLC, $LiNbO_3$, and $SiO_2$ has been measured by diamond indentation using a commercial metrology service (Anton Paar USA Inc.). The result shows that our DLC films are more than twice as hard as conventional soft oxide masks. The different types of films are deposited on the silicon wafers and are etched with the same recipe using the argon ion beam etcher (Veeco Nexus IBE350). By measuring the thickness of the films using a commercial white light spectral reflectometer (Filmetrics F54) before and after etching, we can calculate the etch rate and mask etch selectivity compared to thin film $LiNbO_3$, which is normalized to 1.

**Loss and dispersion measurements.** Loss and dispersion measurements of the ultra-low loss ring resonators (D80_F2_C6_WG1) and the racetrack resonators (D101_F2_C4_WG204) have been performed using frequency-comb assisted diode laser spectroscopy [49] at wavelengths between 1500 nm and 1620 nm.

**Laser frequency noise and tuning measurements:** We performed heterodyne beatnote spectroscopy [66] beating the reference external cavity diode laser (free running Toptica CTL) with the hybrid LNOI laser. The beatnote of the two signals was detected on a photodiode, and its electrical output was then sent to an electrical spectrum analyser (Rohde & Schwarz FSW43). The recorded data for the in-phase and quadrature components of the beatnote were processed by Welch's method [67] to retrieve the single sided phase noise power spectral density $S_{\phi\phi}$ that was converted to frequency noise $S_{ff}$ using: $S_{ff} = f^2 \cdot S_{\phi\phi}$. The frequency noise of the reference laser was determined by a separate beatnote measurement with a commercial ultrastable laser (Menlo ORS), see SI for comparison. We used Tektronix AFG3102 to generate triangular chirp signals with up to 1 MHz frequency. A Keysight DSO204 oscilloscope with 2.0 GHz bandwidth was used to record chirped heterodyne beatnote oscillograms, which were analyzed via a sliding window short time Fourier transform. The length of the Fourier transform window $1/\Delta f$ is optimized for best resolution as the square root of the triangular chirp repetition frequency $T/2$ and the chirp range $B$:

$$\Delta f = \sqrt{\frac{T}{2B}} \tag{1}$$

A Blackman-Harris window function is used to mitigate sidelobes.

**Author Contributions:** Z.L. and R.N.W. fabricated the device with contributions from Z.T.. J.R. and M.C. simulated and designed the devices. J.R., V.S., G.L., and A.S characterized the devices. G.L., J.R., and N.K. performed laser characterization. Z.L., G.L., J.R., and M.C. prepared the figures and wrote the manuscript with input from all authors. T.J.K. supervised the project.

**Funding Information:** This work was supported by Contract HR0011-20-2-0046 (NOVEL) from the Defense Advanced Research Projects Agency (DARPA), Microsystems Technology Office (MTO). Moreover, we acknowledge funding from the European Research Council via grant no. 835329(ExCOM-cCEO) and the Swiss National Science Foundation via the Sinergia grant . J.R. acknowledges funding from the SNSF via an Ambizione Fellowship (No. 201923). M.C. acknowledges funding from the European Union Horizon 2020 Research and Innovation Program under grant agreement No. 812818 (MICROCOMB). A.S. acknowledges funding from European Space Technology Centre through ESA Contract No. 4000135357/21/NL/GLC/my.

**Acknowledgments:** We acknowledge the contribution of Junqiu Liu in the design of testing structures in the early phase of the project. The samples were fabricated in the EPFL center of MicroNanoTechnology (CMi) and the Institute of Physics (IPHYS) cleanroom. The Raman spectrum was measured in the material characterization platform of IPHYS.

**Data Availability:** The code, data and micrographs used to produce the plots within this work will be released on the repository `Zenodo` upon publication of this preprint.